\documentstyle[12pt,aasms4]{article}
\def\ga{\lower 2pt \hbox{$\, \buildrel {\scriptstyle >}\over{\scriptstyle \sim}\,$}}
\def\la{\lower 2pt \hbox{$\, \buildrel {\scriptstyle <}\over{\scriptstyle \sim}\,$}}

\slugcomment{}

\begin{document}

\title{Structure and Kinematics of the Interstellar Medium in front of SN1987A}

\author{Jun Xu and Arlin P.S.~Crotts}
\affil{Department of Astronomy, Columbia University, New York, NY 10027}

\begin{abstract}

High resolution (10~km~s$^{-1}$) [N~II] echelle spectra, sampled every
13~arcsec in a $6\arcmin \times 6\arcmin$ region around SN1987A were obtained
on the CTIO 4m telescope.
The map shows a complicated velocity structure
consistent with that reported previously for the interstellar medium (ISM).
Three components, $V_{hel} =$ 265, 277 and 285~km~s$^{-1}$
were identified as N157C (or the R1170-complex: \cite{xu95}).
The radius of this large superbubble was found to expand at
10~km~s$^{-1}$, with a lifetime of $6 \times 10^6$ years and a total
energy of $3 \times 10^{51}$~ergs determined from its radius and velocity
according to superbubble theory (\cite{mcc87}).
The $V_{hel} = 235$~km~s$^{-1}$ component corresponds to
the near side of 600 pc giant superbubble reported earlier.
This bubble is over $10^7$ years old, and has blown out of the LMC disk.
Two other components, $V_{hel} = $ 255 and 245~km~s$^{-1}$ are identified as
the inner major light echo ring (a double-shell structure) at about 130 pc in
front of SN1987A.
There are also two high velocity components, 300 and 313~km~s$^{-1}$,
which are possibly the far side of a superbubble in which SN1987A exploded.
We also noticed that there are two components at 269 and 301~km~s$^{-1}$
within $20\arcsec$ of SN1987A.
These structures are probably due to the emission from the progenitor star's
red supergiant wind.
We find that the time it took the SN1987A progenitor
to move to the current location 300 pc behind N157C is comparable
to the age of N157C as well as that of the progenitor itself.

\end{abstract}


\section{Introduction}

30 Doradus in the Large Magellanic Cloud (LMC) exhibits extremely
active star formation and interstellar activity.
It serves as an interesting laboratory for the study of the evolution of the
interstellar medium (ISM) and its interaction with massive stars.
Many giant gas sheets in this region were identified in deep $H\alpha$
photographs (\cite{mea80}) and in 21-cm emission (\cite{mea87}).
These shells enclose about ten interlocking superbubbles each about 100~pc in
diameters and expanding at about 30~-~50~km~s$^{-1}$.
Emission and absorption lines (\cite{smi71}; \cite{can80}; \cite{cox83};
\cite{mea84b}; and \cite{cla87}) revealed a complicated velocity structure
which closely correlates with HI components.
The observed components range over $V_{hel}$~=~185~-~307~km~s$^{-1}$.
X-ray maps (\cite{wan91a}, b) demonstrate a good correlation with
$H\alpha$ images, indicating that these superbubbles contain hot gas.
It was shown (\cite{mcc87}) that repeated supernova explosions of massive stars
play the major role in pressurizing these giant bubbles and evacuating
interstellar space, culminating in 600-800~pc diameter supershells.

SN1987A in the LMC provided a unique opportunity for studying part of the
30~Doradus complex.
Ca~II, Na~D and UV absorption lines against SN1987A (\cite{and87},
\cite{vid87}, \cite{mag87}, and \cite{bla88}) exhibit groups of LMC components
near $V_{hel}$~=~129,~171,~206,~226,~250-265 and 281~km~s$^{-1}$.
Other studies (\cite{deb90}) indicate that the components with velocities less
than 140~km~s$^{-1}$ are not associated with the LMC (and are also weaker).

The study of this part of space was greatly
boosted by the discovery of the light echoes from SN1987A (\cite{cro88}).
As forward scattered light from the SN, light echoes carry information not
only about the SN, but also about the interstellar dust which reflect them.
In the decade since the first discovery, these light echoes from SN1987A
have revealed a great deal of information about this region
(\cite{sun88}; \cite{gou88}; \cite{cro89}; \cite{bon89};
\cite{cou90}; \cite{xu94}; and \cite{spy95}).
In a more recent paper (\cite{xu95}), we discuss comprehensively the large
scale ISM structures in the SN1987A foreground, and constructed a
three-dimensional (3D) map of these complicated structures from the light
echoes.
Twelve dust sheets were identified: R310, R430, W700, S730, N980, R1170,
SW1400, NW1500, W1700, R1830, SE3140, and N3240.
These names are derived from the following rules: the first one or two letters
indicate the direction relative to SN1987A, e.g. ``R'' for ring shaped, ``N''
for north and ``SE'' for south-east; the following number is the distance in
front of SN1987A in light years.
These structures enclose a 110~pc radius superbubble which interlocks with a
smaller, 30~pc bubble, a 600~pc diameter giant superbubble, and
a shell which likely envelopes SN1987A.
The giant superbubble appears to have also been observed by another group
(\cite{bru91}) as an oval-shaped dust ring of 1.5-2.0~kpc in dimensions in a
narrow continuum band (4215\AA).

One of the important results from 3D light echo mapping is that the superbubble
N157C is not just the smallest ``cavity'' enclosed by the H$\alpha$ filaments
to the north-east of SN1987A (for example, see \cite{mea80}).
Instead, the complicated H$\alpha$ structure which extends to the south-west of
SN1987A, has a spherical 3D structure whose center of curvature falls on LH90.
Therefore, we think that the N157C supershell extends to a radius of about
115~pc to the south-west, in the direction of SN1987A.
An X-ray map (\cite{wan91a}) reveals a bubble bigger than the small cavity
which was believed to be N157C.
The X-ray maps are likely to show where gas is hitting the inner surface of the
shell, with gas interior to this probably too diffuse to detect.
The lowest contour level (0.0013~photon~s$^{-1}$~arcmin$^{-2}$) of the partial
X-ray ring around the central OB association LH 90 actually extends to a radius
of about 60 pc (4.0 arcmin), but is incomplete to the south and west around its
circumference.
It appears that the N157C bubble extends to larger radii towards the west and
south, towards the direction of the SN.
In the deepest ROSAT PSPC images of the same field (\cite{has96}), it is clear
that two segments of X-ray emitting gas, roughly concentric with respect to the
center of N157C, sit in the region extending into our survey field, the
farthest one at a radius of about 100~pc ($\sim 6.5$~arcmin) with respect to
N157C's center.
Future observations might reveal if these have the same spectrum as the rest of
the X-ray shell.
One cannot prove on the basis of X-ray morphology alone where the edge of the
superbubble extends in the direction of the SN, but it seems likely to be
larger than the 60 pc radius seen at other position angles.
In contrast N157C seen in 21-cm emission from H~I extends over at least 200~pc
in diameter (\cite{kim98}), nearly coextensive with the dust echo.

It is highly desirable to study the velocities of these structures.
Echelle emission spectra in the vicinity of SN1987A were reported before
(\cite{mea90}).
The components were found consistent with absorption structures, with 255 and
280~km~s$^{-1}$ the two most ubiquitous components.
In this paper, we will present the results of our large field echelle survey
around SN1987A and its relation to the ISM structures.
Section 2 discusses observations and data reduction.
Section 3 will explain our numerical model and fitting procedure.
We interpret the velocity map in \S 4.

\section{Observations and Data Reduction}

The survey was conducted on 22-24 January 1993 and on 16-18 January 1994 using
the Cassegrain echelle spectrograph on the CTIO 4m telescope.
We used an echelle setup of five evenly spaced long slits, oriented
north-south.
The slits were $131\arcsec$ long and centered $12\arcsec$.5 apart from each
other.
With such five slits, every exposure covered a field of $131\arcsec \times
50\arcsec$.
By making the offset between fields $120\arcsec$ north-south and $60\arcsec$
east-west (hence $10\arcsec$ between adjacent exposures) there was sufficient
overlap north-south to check that we had returned to the same right ascension
on later nights (while the east-west offsets, also double-checked by
occasionally re-acquiring the SN, were made in quick succession on the same
night).
We planned to use 18 such evenly distributed fields to cover a $370\arcsec
\times 350\arcsec$ region, from $125\arcsec$ south to $245\arcsec$ north of
SN1987A, and from $170\arcsec$ west of to $180\arcsec$ east of SN1987A.
We managed to get 17 fields, not having enough time to observe the
south-westernmost corner field.
Additionally, two single $247\arcsec$-long slits oriented east-west, one
centered on the SN and one centered $100\arcsec$ north of the SN, were also
used beforehand to check for structure that might extend beyond the
120~km~s$^{-1}$ free velocity range between adjacent slits.
No such structure was found, with the exception of a small region in the
five-slit data at the edge of the field southeast from the SN.
This corresponds to the ``Honeycomb Nebula'' (\cite{wan92}, \cite{mea93},
\cite{mea95}, \cite{chu95}), where we ignored some data which appeared
susceptible to overlap.
The configuration of all of these slit positions is shown in Figure 1.

We chose to look at the [N~II] doublet (air $\lambda 6548.08$ and $\lambda
6583.41$: \cite{ost92}), and used an H$\alpha$ filter (central
wavelength~=~6563\AA, FWHM=75\AA) to separate orders.
We use [N~II] instead of H$\alpha$ because the thermal width of H$\alpha$ line
is about 3.7 times of that of [N~II] line at the same temperature due to the
difference in their atomic mass, a critical difference.
Resolution of the setup is 10~km~s$^{-1}$ near [N~II] $\lambda 6583.41$,
and the free velocity range between every two slits is 120~km~s$^{-1}$.
Exposure time of each field is 3600 seconds.

We used IRAF to reduce the data.
The process started with column bias, two-dimensional bias frame subtraction,
and dark current subtraction.
Bad pixels were interpolated over.
The response function image from each night, which was later used to flat field
the images, was constructed from combined quartz exposures using the IRAF
``response'' task.
We then used the IRAF ``illumination" task to fit ``slit functions'' from sky
exposures.
These slit functions were then applied in illumination correction to all the
object images to give a flat response along slits.
We calibrated the spectra into wavelength space using ThAr comparison spectra.
The calibration error can be assessed by comparing the measured sky line
wavelength to the standard values (\cite{hub86}).
We found the error was less than 0.04 \AA, or equivalently 2~km~s$^{-1}$ at
$\lambda 6548$.
This error is only 1/5 of the 10~km~s$^{-1}$ resolution.
Because the later process requires a high S/N ratio, we binned the spectra over
every $13\arcsec.3$. 
A continuum was then fit and subtracted from each one of the $13\arcsec.3$
spectra.
We then scaled the spectra in adjacent fields to the same flux level.
The scaling was possible because we deliberately overlapped each field with its
neighbors.
We multiplied spectra of each field by a constant so that neighboring fields
have the same $H\alpha$ and [N~II] ($\lambda 6583$) flux in the common region.
Finally, night sky lines were properly subtracted from the spectra.
Unfortunately, one of the [N~II] doublet ($\lambda 6548.08$) was terribly
contaminated by OH-lines.
Therefore, we only look at the other [N~II] line $\lambda 6583.41$.

\section{Velocity Components Analysis}

We transformed the spectra into velocity space using the Doppler shift formula. The [N~II] vacuum wavelength are $\lambda 6549.87$ and $\lambda 6585.21$ which
are calculated from their air wavelengths with Edlen's formula (\cite{edl66}).

\subsection{Component Identification}

It is crucial to first identify structures corresponding to each slit,
because the 5-slit echelle may cause cross-slit overlap.
For this purpose, we looked at two single slit spectra:
one centered on SN1987A, and the other about $100 \arcsec$ to the north.
We first binned them over $130\arcsec$ to achieve a high S/N level.
Figure 2 show H$\alpha$ and [N~II]$\lambda 6583$ lines of these spectra.
In Panels 1 and 3 (H$\alpha$ line), the two structures
near 0 and 650~km~s$^{-1}$ are skylines $\lambda 6562.67$
and $\lambda 6577.21$ (\cite{hub86}) respectively.
In Panels 2 and 4 ([N~II] $\lambda 6583$ line), the structure
near 600~km~s$^{-1}$ is a skyline $\lambda 6596.55$ (\cite{hub86}).
These spectra show that the only interesting structures are in
210-330~km~s$^{-1}$ interval, which is within our slit-to-slit free range.
Although this conclusion is technically correct only in the region on and to
the north of SN1987A, we think it is very unlikely that
some extraordinary low or high velocity structure would appear only to the
south of SN1987A, since most structures are seen to be weaker there.

\subsection{Numerical Analysis}

The goal of numerical analysis is to identify the number of velocity components
and to measure the properties of each component, e.g.~velocity, intensity and
FWHM.
Our analysis was a two-step process.
We first fit the data with various numerical models with up to five gaussian
components.
Then based on the fitting result, we used a statistical algorithm to select the
correct component numbers.

First, we varied the number of Gaussian components from one to five to produce
five different models.
These models were fitted separately to the spectra using least $\chi^2$
fitting to determine the parameters.
Because the temperature of each component in the LMC should not change very
much, and hence the FWHM of each component should be roughly the same, we
forced a uniform FWHM across components within each model, while the height and
position were allowed to vary freely and independently from each other.
To check out the ``uniform FWHM'' assumption, we repeated the whole process but
let FWHMs vary as free variables.
The ``model probability'' method (see below) selected the same number of
components in every field, regardless the assumption on FWHMs.
We also noticed that the velocities of corresponding components are a little
different, but within our resolution.
Therefore, we think our modeling based on the ``uniform FWHM'' assumption is
robust.
The fitting was performed with the ``deblend'' option of IRAF's ``splot'' task. We chose to work interactively because the components are normally not well
separated and automatic fitting tends to yield wild results.

Based on the least $\chi^2$ fitting result, we computed the probability that
a ``correct'' model would yield a higher $\chi^2$ than the actually fitted
result. 
The selection rule is that the most likely model has the maximum probability.
It has been shown that the probability defined in this way satisfies a
$\chi^2$-distribution and consequently can be calculated  (\cite{pre92}).
For every profile, we compared the five models based on their probabilities,
and select the one with the highest probability.
In most cases, we found the ``correct'' model yields a probability several
orders of magnitude higher than other models, and hence convincingly
establishes itself.
For a few profiles, two or three models have similar probabilities.
In these situations, we made our choices by continuity argument.
We look at the neighboring locations for any definitely
``correct'' models, and select the most similar one. 
Figure 3 shows the reduced data from one of our 17 fields, detailing the 
[N~II]$\lambda 6583$ line, and the result from our velocity component
analysis graphed on a corresponding scale in the adjacent diagram.
The analysis reveals all of the major features and nearly all weaker ones,
and follows gradients in the velocities of components with reasonable
accuracy.

With the aforementioned process, we determined at every
grid point over all 17 fields the number of components and properties of
each component.
We then looked into the connection among these grid
points for collective structures across the field.
The result will be discussed in the next section.

\section{ISM in the Foreground of SN1987A}

This survey shows the following ionized clouds: $V_{hel}$ = 237, 245, 255, 265,
277, 285, 300, and 313~km~s$^{-1}$.
(When clouds are within about 5~km~s$^{-1}$ of each other, they tend to blend
together.)~
These are shown in Figure 4a.
The $265-285~$km~s$^{-1}$ grouping is sufficiently complex that we show it
separately in Figure 4b.
Four of these components (255, 277, 300 and 313) were
also reported in a smaller field containing SN1987A (\cite{mea90}).
Two of them, $V_{hel}=255$ and 277~km~s$^{-1}$ were discovered in the
interstellar absorption spectra toward SN1987A (see the citations in \S 1).
A HI 21-cm study in this region exhibited similar profiles:
$V_{hel}$ = 243, 274, 296 and 318~km~s$^{-1}$ (\cite{mea84a}).
Similar structures were also discovered in the halo of 30
Doradus from [O~III] lines (\cite{mea84b}; \cite{mea88}).
In our survey, we found the 245 and 255~km~s$^{-1}$ components (Group B in
Figure 4a) have similar properties and hence are likely to be two layers of one
structure (see \S 4.3 below).
Similarly, $V_{hel}$ = 265, 277 and 285~km~s$^{-1}$ clouds (Group C and Figure
4b) are seen to be interrelated with each other and even connected in some
places.
(Some components are seen to disappear into others when approach within the
5~km~s$^{-1}$ blending range.)~
We think these three are really one structure (see \S 4.1 below).
The two high velocity clouds (300 and 313~km~s$^{-1}$; Group D) are likely to
be the far side of the 245-255~km~s$^{-1}$ shell (\S 4.3, see also
\cite{mea90}).
The other component (237~km~s$^{-1}$; Group A) is reported for the first time
in this study.
It is interesting that we did not see this component in the two single long
slit spectra, one centered on SN1987A and the other $100 \arcsec$ north of the
supernova.
We show in \S 4.2 that this component correlates to the two clouds
about 1 kpc in front of SN1987A (\cite{xu94}).

\subsection{N157C}

The $V_{hel}$ = 265, 277 and 285~km~s$^{-1}$ group (C in Figure 4a, also seen
in Figure 4b) is the most far reaching structure in this field.
They not only are interrelated with each other, but also have comparable flux
level consistently bright throughout the whole field.
Similar behavior was observed in [O~III] and
$H\alpha$ emission lines (\cite{mea90}), in the HI 21-cm survey
(\cite{mea84a}), and in Ca~II, Na~D and UV absorption lines (see \S 1 for
references).
Therefore, it is tempting to recognize these
three components as one dominant structure in front of SN1987A.
In an earlier paper (\cite{xu95}), we show that the most
prominent structure in the SN foreground is the echo 1170 light-years in the
SN foreground, the R1170-complex (identified as N157C).
This similarity marks the first correlation
indicating that these velocity components are indeed N157C.

As another piece of evidence, we plotted the flux contour of these components
on top of the 2D image of the ISM dust structure, the
R1170-complex (\cite{xu95}) (Figure 5).
This plot shows a general tendency that both structures are bright
to the north of SN1987A.
In addition, we observed a good correlation not only on the large scale but
also at many small bright spots, limited by the spatial accuracy of the echelle
survey of $13\arcsec$ east-west and $40\arcsec$ north-south.

We also observed that the two structures split in the same way.
The 3D light echo map (\cite{xu95}) demonstrated that the cloud, consisting of
N1500 and W1700, splits from R1170 to the north of SN1987A
and joins R1830 southwest of the SN.
Similarly, Figure 4b shows the 277~km~s$^{-1}$ component joining the
285~km~s$^{-1}$ component to the south (and slightly east) of SN1987A and the
265~km~s$^{-1}$ component to the southwest.
This geometry is consistent with the picture that the 285 and the
265~km~s$^{-1}$ components are the far and near shell of N157C, while the
277~km~s$^{-1}$ component is a cloud joining the two structures to form a
superbubble.

N157C, with a diameter of about 200 pc, is expanding at a speed of
10~km~s$^{-1}$, therefore.
We believe this bubble is still in the pressure-driven
stage for the following reasons.
First, this size is common in other parts of the LMC,
particularly in 30 Doradus (\cite{mea80}).
Most of these giant bubbles are young and rapidly expanding (\cite{mea87}).
McCray and Kafatos (1987) attributed the generally large superbubble
size to the thick gas disk of the LMC and its low metallicity.
Secondly, the interior of a pressure-driven bubble is filled with hot gas as
opposed to the cool interior in the snow-plow stage.
In addition, Wang and Helfand (1991b) reported X-ray ($\approx 0.8 - 1.2 keV$)
emission in N157C, corresponding to very hot gas in its interior.
Therefore, N157C is in the energy-conservation stage.
McCray and Kafatos (1987) gave the radius and velocity properties for these
bubbles:
\begin{eqnarray}
R_{S} = 97~ pc~ (N_\star E_{51} / n_{0})^{1/5}t_7^{3/5} \\
V_{S} = 5.7~km~s^{-1}\: (N_\star E_{51} / n_{0})^{1/5}t_7^{-2/5}
\end{eqnarray}
Dividing one equation by the other,
\begin{eqnarray}
t_7 = \frac{R_{S}}{V_{S}} \frac{5.7~km~s^{-1}}{97~pc}
\end{eqnarray}
where $n_0$ is the gas atomic density, and $t_7 = t / 10^7~$yr.
Let $R_S = 100$~pc, $V_S = 10$~km~s$^{-1}$.
We estimate the age of N157C to be $t = 6$~Myr.
Therefore, this superbubble is still very young.
It is estimated (\cite{mcc87}) that stellar winds from a rich cluster
($N_\star \geq 10$) dominate its superbubble expansion for a few million
years until supernova explosions become more important, while the superbubble
may expand to less than $\approx 100$ pc.
We think N157C has just passed the stellar-wind-driven stage, and is in rapid
expansion driven by powerful supernovae from LH90.

From other sources, we estimate the neutral hydrogen column density $N_{HI}$ of
the supershell N157C.
From 21~cm emission (\cite{luk92}), about 70\% of the H~I column density 
towards LH90 is contained in a narrow component centered at 270~km~s$^{-1}$,
with a total column density of all components of about
$2.2\times10^{21}$~cm$^{-2}$.
This total $N_{HI}$ is corroborated from the pointing (\#43) containing N157C
from the survey of \cite{mea87}: $N_{HI}=2.3\times10^{21}$~cm$^{-2}$, and the
value inferred from the damping wings of the Ly $\alpha$ absorption line
towards the star Sk~$-69^\circ203$, in the direction of N157C:
$2.4\times10^{21}$~cm$^{-2}$ (\cite{fit90}).
Thus the value of $N_{HI}$ for N157C is about $1.5\times10^{21}$~cm$^{-2}$.
(However, none of these surveys have resolution better than about $15\arcmin$,
so we cannot be sure of the effects of substructure.)~
Since this column density is optically thick at the Lyman limit, we will assume
the gas is shielded and therefore largely neutral.
Accounting for projection effects (a factor of four), the kinetic energy of the
expanding HI shell is therefore $E_{shell} = 5.4 \times 10^{50}$ ergs.
This implies a total kinetic output from SNe that is $77/15$ times greater
(\cite{mac88}): $E_S = 2.7\times10^{51}$~ergs.
Each typical Type II supernova of a massive O star releases about
$E_{SN} = 10^{51}$ ergs of kinetic energy.
Given an average supernova rate of about one per million years in a typical OB
association (\cite{mac88}), the energy budget is consistent with the
aforementioned age estimate.

Given $E_{shell}$ determined from theoretical consideration of a SN driven
expansion (\cite{mcc87}):
\begin{eqnarray}
E_{shell} = 4.0 \times 10^{49} ergs~ (N_\star E_{51}) t_7
\end{eqnarray}
we find $N_\star E_{51} = 22$.
Combining this with Equation 1 (or 2) yields an initial density $n_0 = 4$.
This is a reasonable estimate, corresponding to the
$N_{HI} = 2.4\times10^{21}$ cm$^{-2}$ spread over a scale height of $100$~pc,
which is reasonable in terms of the observed 3-D structure.

On the other hand, from stellar content point of view, the aforementioned
estimate $N_\star E_{51}$ implies several massive SN explosions in LH90,
given that the typical energy of a Type II SN, $E_{51}$,
is a few for the most massive stars ($M \ge 30 M_{\sun}$).
The main-sequence lifetimes of massive stars are given approximately by
$t_{MS} \approx 30$~Myr$ \; (M_\star / [10 M_{\sun}])^{-\alpha} $,
where $\alpha \approx 1.6$ for $7 \leq M_\star \leq 30 M_\star$
(\cite{sto72}) and by $t_{MS} \approx 9$~Myr$ \;
(M_\star / [10 M_{\sun}]^{-0.5}$ for $30 \leq M_\star \leq 80 M_\star$
(\cite{chi78}).
In a time scale of 6~Myr, only a few very massive stars with
($M_\star \ga 30 M_{\sun})$ would explode.
It is estimated that an OB association should produce roughly 9 times as many
stars with masses in the range 7-30 $M_{\sun}$ (main-sequence spectral type
B3-08, corresponding to $N_\star$) as stars with masses $\geq 30 M_{\sun}$
(MS type O7-03) (\cite{mcc87}).
Given that there are about 50 stars with mass $\ge 7 M_{\sun}$ (\cite{hil94}),
there should be about 5 massive stars of $M_\star \ga 30 M_{\sun}$ which
should have exploded.

We can also compare our predicted SN explosion number to
empirical determinations of the stellar population.
In addition to the original study of LH 90 (\cite{luc70}), two recent studies
of the stellar population of the association yield somewhat different results.
Recent UIT data (\cite{hil94}) on LH90 and six other LH OB associations in all
but the inner portion of the 30 Dor complex give the number of stars and their
inferred mass.
For the clusters with the largest mass stars ($\approx 50 M_\odot$), the ratio
of $M \geq 30 M_\odot$ versus $7 M_\odot \geq M \la 30 M_\odot$ stars predicted
in the previous paragraph from theoretical estimations appears inconsistent
with the UIT data, in the sense that more massive stars are actually seen by
UIT.
From these data, LH90 appears to have stars as massive as $41 M_\odot$, while
LH89 and LH99 show about eight stars more massive than this (out of a total of
40 stars in LH90 and 68 stars in LH89 and LH99 more massive than $15 M_\odot$).
Thus LH90 might be missing approximately five stars of $M\approx 50M_\odot$ (or
more if stars have exploded in LH89 or LH99); these have presumably formed SNe.

We should note, however, the existence of stars in LH90 more massive than those
allowed ($\ga 30 M_{\odot}$) by a 6~Myr year turnoff age.
Perhaps some massive stars were created millions of years after the first
stars.
Indeed, our prediction for the number of massive stars seems more consistent
with the results from the other study (\cite{tes93}), which uses optical
photometry and spectroscopy of LH 90 to derive spectral types and $M_{bol}$
values consistent with about 7 stars with $M_\star \ga 30 M_{\sun}$ versus
more than 47 stars with mass $\ge 7 M_{\sun}$ (according to our simple
conversion of $M_{bol}$ to $M_\star$).
Indeed, their study concludes that the association contains a sub-population
3-4~Myr old and a possible larger sub-population 7-8~Myr old (\cite{tes93}).
This is consistent with our arguments regarding the SN rate.

\subsection{600 pc Giant Superbubble}

Xu, Crotts and Kunkel (1994) first reported two dust bodies at
about 1 kpc in front of SN1987A.
They (\cite{xu95}) also found that the southeast piece of these two correlates
with one of the bright $H\alpha$ filaments extending from N157C discovered by
Meaburn (1984).
They went one step further to suggest this filament as the near side and
N157C as the far side of a giant superbubble of 600 pc in diameter.
Consistent with this picture, a component at 235~km~s$^{-1}$ (``A'' in Figure
4a) has been found in the regions far away from SN1987A.
Figure 6 plots flux contour of this component
over the image of SE3140 and N3240 (\cite{xu95}).
We found that one piece of the 235~km~s$^{-1}$ components
corresponds exactly to N3240, and another piece correlates with SE3140.
It is unfortunate that we only had enough time
to finish the survey to the north of Dec $-69^\circ 18\arcmin$
and to the west of RA 5h 36m 00s (in J2000) delineated by the dashed lines.
Still, the available data indicates that 235~km~s$^{-1}$ gives a good
correlation with SE3140, expanding at about 40~km~s$^{-1}$ relative to the
centroid of N157C.

It is interesting to see the 235~km~s$^{-1}$ component distributed in a patchy
ring-like shape.
Recent light echo images (1995-1997) showed N3140 becoming larger and
extending to the position angle $PA = 45^\circ$
where the $(5h36m, -69^\circ 14\arcmin)$ 235~km~s$^{-1}$ components is located.
Therefore, we think the 1 kpc ISM structure may indeed show a hole in front
of SN1987A.
Moreover, we have several reasons to believe this hole may relate to the
blow-out of the 600-pc-diameter superbubble.
First, we have not observed any structure within $3\arcmin$ of SN1987A either
in the light echoes at 1 kpc or in the echelle survey at 235~km~s$^{-1}$,
indicating that the cloud has been fragmented.
This result can be explained by a superbubble blow-out.
Secondly, Wang and Helfand (1991b, c) reported that this region has soft
X-ray emission much higher than the background but much lower than other young
superbubbles in 30 Doradus.
This region may even be contained in the possible giant superbubble LMC-2,
roughly 1~kpc across (\cite{wan91c}) which has blown out of the disk.
Finally, 600 pc is one of the largest superbubbles ever observed in the LMC
(\cite{mea80}), and therefore may represent a size close to the upper limit
imposed on the superbubble size by the LMC disk thickness.
In conclusion, we think this 600 pc diameter giant superbubble appears to have
blown out of the LMC disk and is expanding in radius (assuming it is pushing
on one side against N157C) at 20~km~s$^{-1}$.
McCray and Kafatos (1987) suggested that when the superbubble cools or
breaks out of the galaxy, it will be distorted by gravitational instability,
collapse and form giant clouds, and finally ignite propagating star formation.
LH90, which formed N157C, may have itself formed in this way in the
gravitational collapse of the 600 pc giant bubble.

If this superbubble blew out in its snow-plow stage, prior to this it was
described in terms of the time ($t_c$) and expansion radius ($R_{c}$) beyond
which cooling becomes important in the interior e.g.~post-adiabatically
according to (\cite{mcc87}):
\begin{eqnarray}
R(t) \approx R_c (t / t_c)^{1/4}
\end{eqnarray}
for
\begin{eqnarray}
R_c = 50~ pc \; \zeta^{-0.9} (N_\star E_{51})^{0.4}~ n_{0}^{-0.6}
\end{eqnarray}
and 
\begin{eqnarray}
t_c = 4\times 10^6 \; \zeta^{-1.5}( N_\star E_{51})^{0.3}~ n_{0}^{-0.7}
\end{eqnarray}
where $\zeta$ is metallicity, $N_\star$ is number of stars $\geq 7M_{\sun}$,
$E_{51} = E / 10^{51} ergs$, $n_0$ is gas atomic density.
In the LMC, $\zeta\approx0.3$ (\cite{duf84}) and a density typical for the LMC
(several times lower than near N157C), $n_0\approx0.35$ (\cite{hin67}).
Let's guess $N_\star = 50$ (roughly that of N157C) and $E_{51} =
1$, hence $t_c=160$~Myr, and $R_c\approx 1.3$~kpc, much bigger than 300~pc.
Therefore, we think this giant bubble was probably hot when it blew out,
implying an age at the time of blowout of about 10~Myr from eq.~3, or 15~Myr
using eq.~1 and the (less certain) parameters above.
These are probably consistent with the age of LMC-2 (\cite{wan91c}).
This appears older than N157C, indicating the 600~pc superbubble predates it,
and may have even brought about LH90's formation.

\subsection{Interstellar Environment of SN1987A}

\cite{xu95} demonstrated that the two dense clouds
(R430-complex including R310) about 100~pc in front of SN1987A do not
correspond to any prominent $H\alpha$ structures, and therefore are likely 
related primarily to SN1987A.
R430 is brighter and larger than R310.
The lack of correlation is also consistent with the assertion that SN1987A
exploded about 300~pc on the far side of the LMC disk (\cite{xu95}).
Figure 7 plots flux contour of $V_{hel}$ = 255 and 245~km~s$^{-1}$ (Group B in
Figure 4a) on top of R430-complex, indicating a good correlation between these
two images.
On the other hand, the 255~km~s$^{-1}$ component is brighter and larger than
245~km~s$^{-1}$ component, in good agreement with the relative size and
brightness of R430 and R310.
The correlation between these two studies indicates that the double shell
structure R430 and R310 have velocities 255~km~s$^{-1}$ and 245~km~s$^{-1}$
respectively.
The 255~km~s$^{-1}$ component was reported by other groups
before (see \cite{mea90} and the reference therein).
However, Meaburn (1990) suggested that the 255~km~s$^{-1}$
shell is within 10 pc of SN1987A.

Two other ionized clouds $V_{hel}=300$ and $313~km~s^{-1}$ in the echelle
spectra (Group D in Figure 4a) do not correspond to any absorption structures.
It was suggested (\cite{mea90}) that these two clouds are likely to be the far
side of a superbubble which envelops SN1987A.
We noticed that this identification is consistent with the
double shell structure on the near side.
Unfortunately, the light echoes have expanded to only
several light years behind SN1987A.
So far, we do not have any direct light echo results on these two possible
double shells on the far side of the supernova.
Therefore, we can only speculate on the structures
of these two high speed components.

If these two components are indeed on the far side of a superbubble surrounding
SN1987A, this superbubble is expanding in radius at 30~km~s$^{-1}$.
If we adopt a average radius of 90 pc, the age of this bubble is about 2~Myr,
(eq.~1).
Let the shell density of this bubble $N_{H} \approx 1 \times 10^{20}$
(\cite{mea87}), using the same method in \S 4.1, we calculate the total
energy of this superbubble to be $3 \times 10^{51}$ ergs, which is the energy
of two to three typical Type II supernovae.
Both the energy budget and the age suggest that there had been a small number
of supernova explosions before SN1987A in this superbubble.
Because of the short lifetime of this bubble, these previously exploded
supernovae must be very massive stars ($M_\star \ge 30 M_{\sun}$)
(\cite{chi78}).

The aforementioned superbubble assertion has some problems.
First, there are not many stars close to SN1987A.
Even though one may argue that massive stars might have exploded, the
lack of low mass stars is still a difficulty.
Second, the expansion velocity of 30~km~s$^{-1}$ is too high compared to that
of N157C (10~km~s$^{-1}$) which is pressurized by the second richest cluster
LH90.
The only way around this problem is to assume very low gas density in this
region.
In Equation 2, adopt $V_S = 30$~km~s$^{-1}$, $N_\star E_{51} = 3$,
$t_7 = 0.2$, we get $n_0 = 0.02$, which is an order of magnitude lower than
the typical LMC value of $n_0 = 0.35 $ (\cite{hin67}).
Perhaps this is reasonable, given SN1987A's likely distance from the plane of
the LMC, but the existence of a superbubble surrounding SN1987A is still not
certain.

We also noticed that there are two components at 269 and 301~km~s$^{-1}$ within
$20\arcsec$ of SN1987A.
They have much higher flux level than those probably more distant components at
245, 255 and 317~km~s$^{-1}$ (and at 300~km~s$^{-1}$ not within $20\arcsec$).
It has been suggested that these structures are probably
due to the emission from the red supergiant wind (\cite{crku91}).
This structure will be treated in a separate paper (\cite{cro98}).
These two components have been reported previously (\cite{mea90}).

\section{Discussion}

We note that our light echo 3-D map, in conjunction with the velocities of the
matter in front of the SN from this study, implies that the SN is well behind
most of the material in its portion of the LMC, and hints at how this occurred.
This refutes claims made on the basis of extinction towards a nearby star,
Sk~$-69^\circ$203, that the SN is embedded in the middle of the main H~I
complex (\cite{fit90}).
We approximate the speed of LH90 with that of N157C, about 270~km~s$^{-1}$.
This should also represent the speed of the LMC stellar disk.
Therefore, SN1987A is moving away from the LMC disk at a speed of
20~km~s$^{-1}$ given that the velocity of SN1987A of 289~km~s$^{-1}$
(\cite{cro91}).
At this speed, it would have taken the SN1987A progenitor $\sim 10^7$ years
to move to its current location 300~pc away from the stellar disk.
It is interesting to compare this time scale to the lifetime
of N157C (6~Myr, \S 4.1) as well as that of the speculated superbubble
surrounding SN1987A (2~Myr, \S 4.3), and that of the progenitor itself.
We speculate that the SN and N157C were created together (followed by the
superbubble around SN~1987A), and are pursuing further methods to test this.

The authors would like to thank Cerro Tololo Inter-American Observatory for
granting observing time to this study and all the wonderful support that we
have received.
We would also like to direct our gratitude to Rob Olling and David Helfand for
many enlightening discussions.
This research was supported by NSF grant AST 90-22586, the A.C.'s David \&
Lucile Packard Foundation fellowship, and NASA LTSA grant NAG5-3502.

\newpage

\begin{center}
\bf \large Figure Legends
\end{center}
\rm \normalsize
\vspace{0.5 cm}

Figure 1: A 15$\arcmin$ field centered on SN 1987A imaged in a band
highlighting H$\alpha$ and [N~II], and showing the positions of our echelle
long slits relative to features in the field.
The OB association LH90 sits near the corresponding label, with the most
prominent groups of stars seen to the west and south of the label.
It is surrounded at radii out to $\sim 10\arcmin$ by emission from gas
associated with the accompanying supershell N157C.
The Honeycomb Nebula (labeled ``HN'') sits about 2.5$\arcmin$ southeast of
the SN.
At the time of these observations, the light echoes extended to roughly the
edge of our slit array.

Figure 2: Binned spectra of two single $120\arcsec$ long slits.
From upperleft counterclockwise are Panels 1, 2, 3 and 4.
Panel 1: $H\alpha$ spectrum centered on SN1987A;
Panel 2: [N~II] ($\lambda 6583 $) spectrum centered on SN1987A;
Panel 3: $H\alpha$ spectrum $100\arcsec$ north of SN1987A;
Panel 4: [N~II] ($\lambda 6583 $) spectrum $100\arcsec$ north of SN1987A.
The abscissa is $V_{hel}$ in km~s$^{-1}$.
The ordinates are in counts of each pixel.
In Panels 1 and 3, the structures near 0 and 650~km~s$^{-1}$ are night sky
lines.
In Panels 2 and 4, the structures at 600~km~s$^{-1}$ are also night sky lines.

Figure 3a: The reduced data from one of our 17 subfields (the one just
northwest of the SN).
The range of data is 120~arcsec in the vertical direction (along the slit),
with a free velocity range of 120~km~s$^{-1}$ corresponding to each of the five
slits in the horizontal direction.
The SN itself is seen in [N~II]$\lambda$6583 in the lower (southern) left
corner.

Figure 3b: On the same scale as Figure 3a, the inferred velocity components are
shown for the same data.
The emission strength of each component (slit into 13.3~arcsec subsamples) is
indicated by the width of each line sgment, which varies with the logarithm of
linestrength.
The tickmarks along the vertical axis indicate divisions of 13.3~arcsec and
those along the horizontal indicate steps of 50~km~s$^{-1}$ (in the frame of a
single slit).

Figure 4a: 3-D surface plot of all eight velocity components in front of
SN1987A.
They are collected into four groups, A, B, C and D (described in \S 4), with
labels of each letter at the corners of each group indicating its approximate
extent on the plot.
The vertical axis is $V_{hel}$ in km~s$^{-1}$.
The other two axis are RA and Dec in J2000.
North is to the left.
West is the direction pointing into the paper.
Note that structure within about 5~km~s$^{-1}$ will blend by one of them
disappearing.

Figure 4b: 3-D surface plot of the three velocity components in Group C,
265, 277 and 285~km~s$^{-1}$.
The axes and orientation are the same as in Figure 4a.
Notice that the 277~km~s$^{-1}$ component joins the 265~km~s$^{-1}$ component
to the southeast of SN1987A, and the 285~km~s$^{-1}$ component to the south and
slightly east of the SN.

Figure 5: Flux contour of combined $V_{hel} = $ 265, 277 and 285~km~s$^{-1}$
components on top of the R1170-complex image (\cite{xu95}).
The axes are RA and Dec (J2000).
Both the flux- and the dust-maps are smoothed in both directions over
$13 \arcsec$, which is the resolution limit of the echelle survey.
This survey only covered the region within the dashed lines.
The ``holes'' in the gray scale dust-map are due to the lack of data and
inadequate contrast in image display after smoothing.
The readers are encouraged to compare this plot and Figures 6 and 7 with the
similar plot in \cite{xu95}.

Figure 6: Flux contour of the $V_{hel} = $ 235~km~s$^{-1}$ components
plotted over the combined SE3140 and N3240 signals in greyscale (\cite{xu95}),
analogously to Figure 5.

Figure 7: As in Figure 5 and 6, flux contour of combined $V_{hel} = $ 245 and
255~km~s$^{-1}$ components plotted over greyscale image of the R430-complex
(\cite{xu95}).

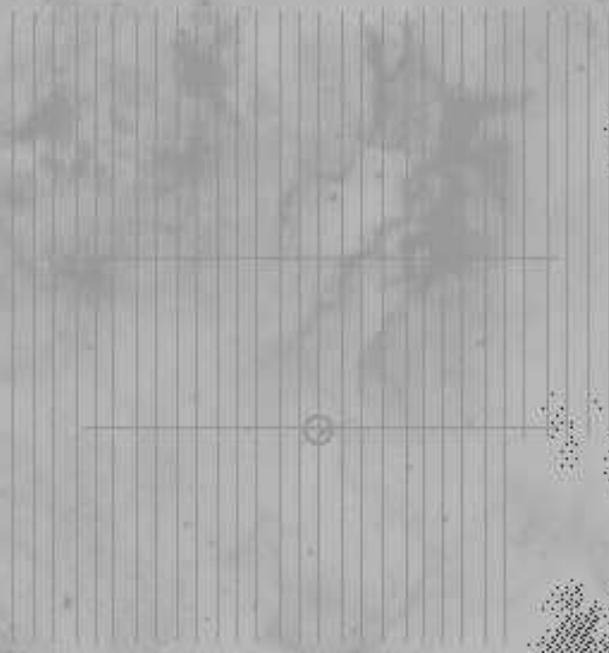
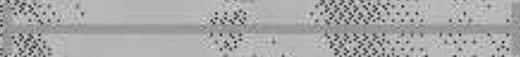
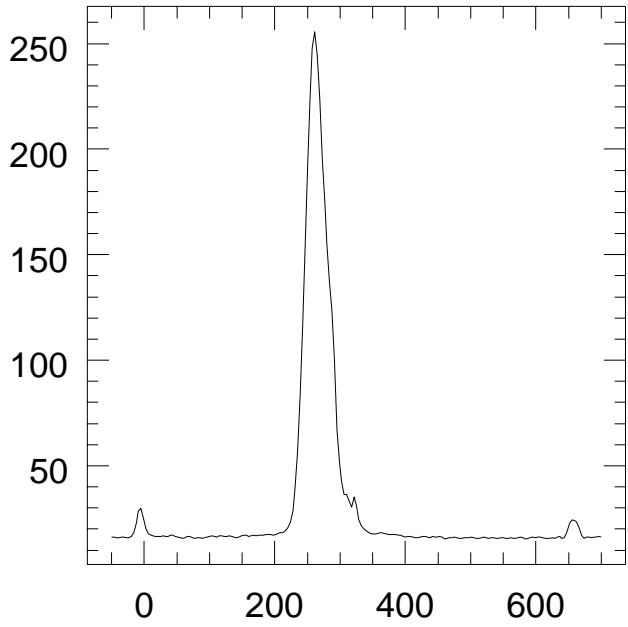
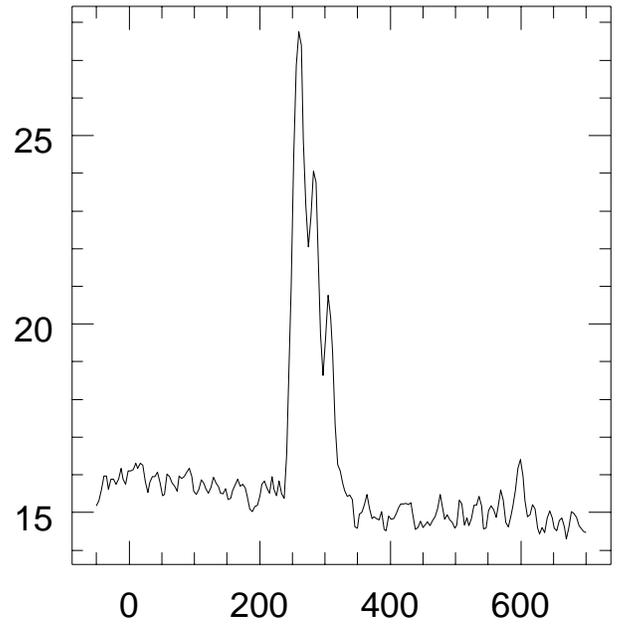
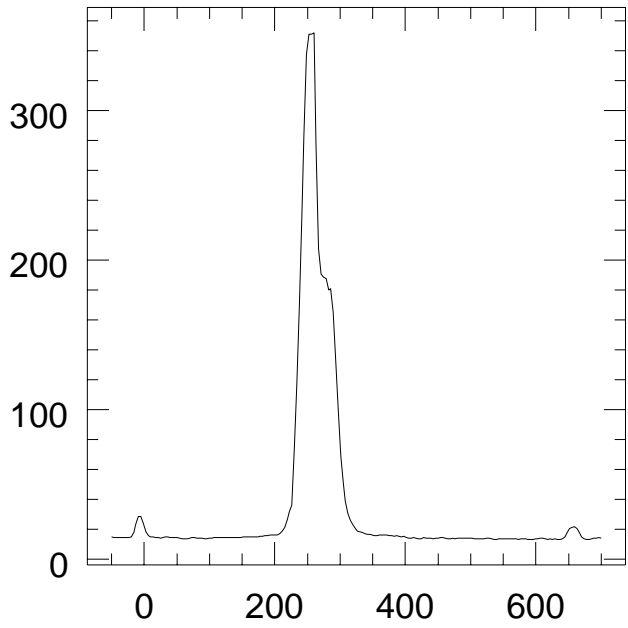
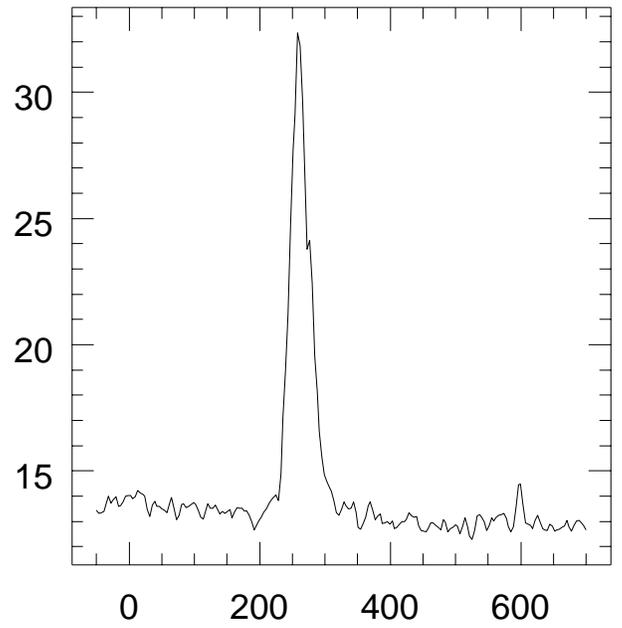
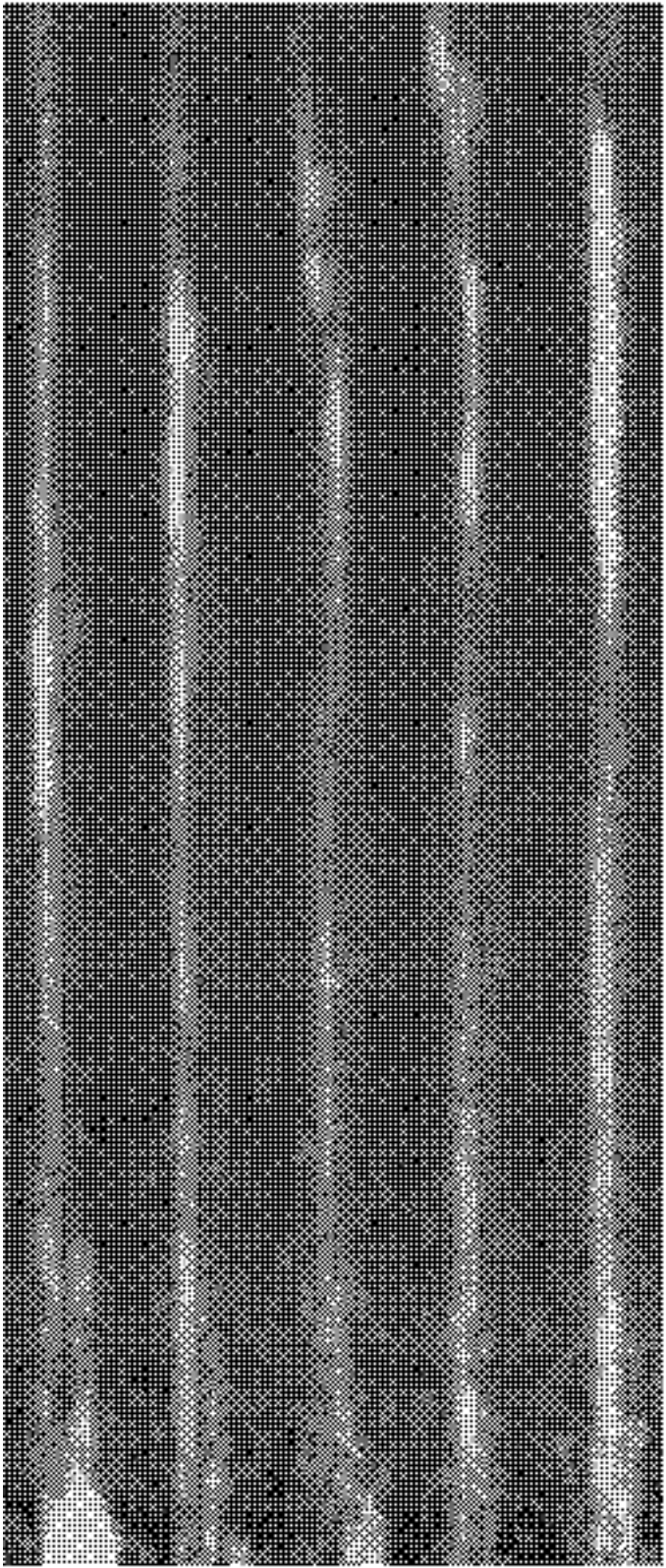
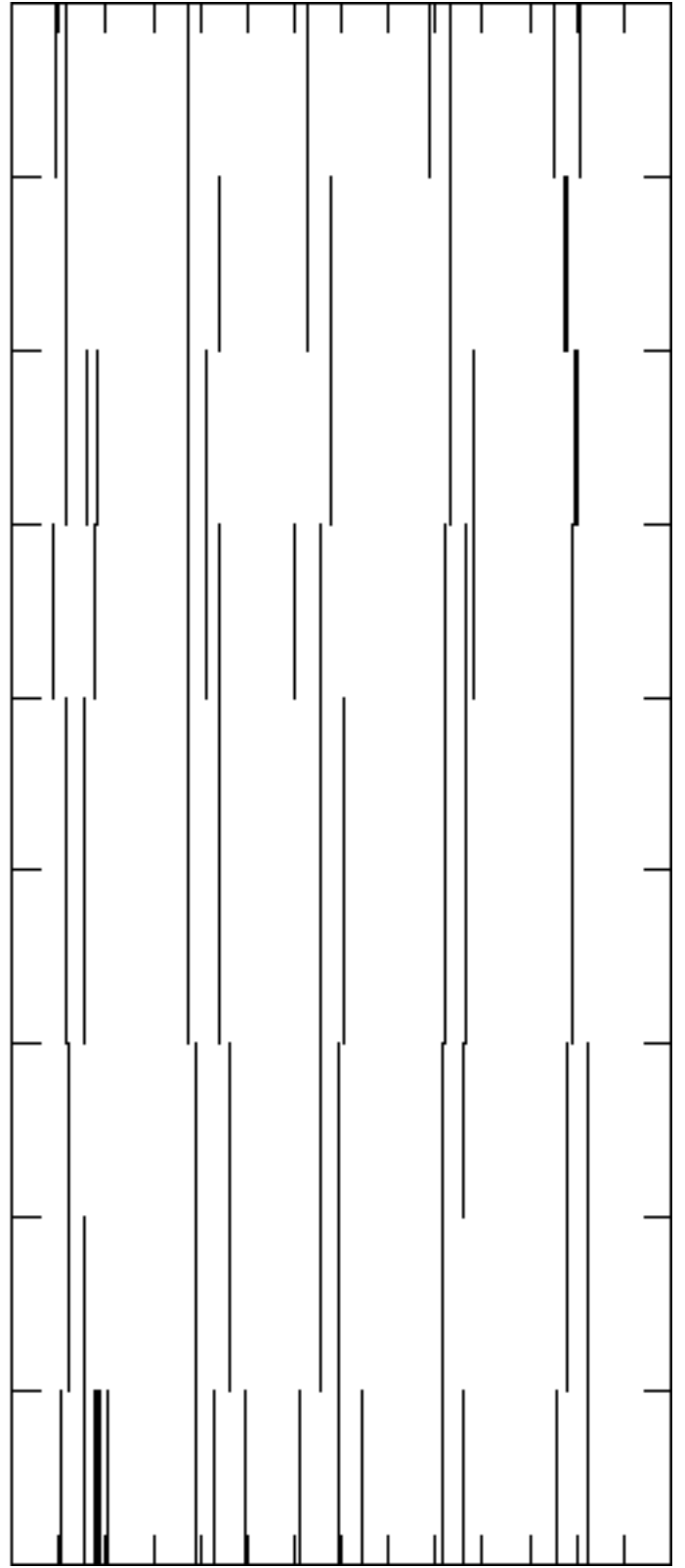
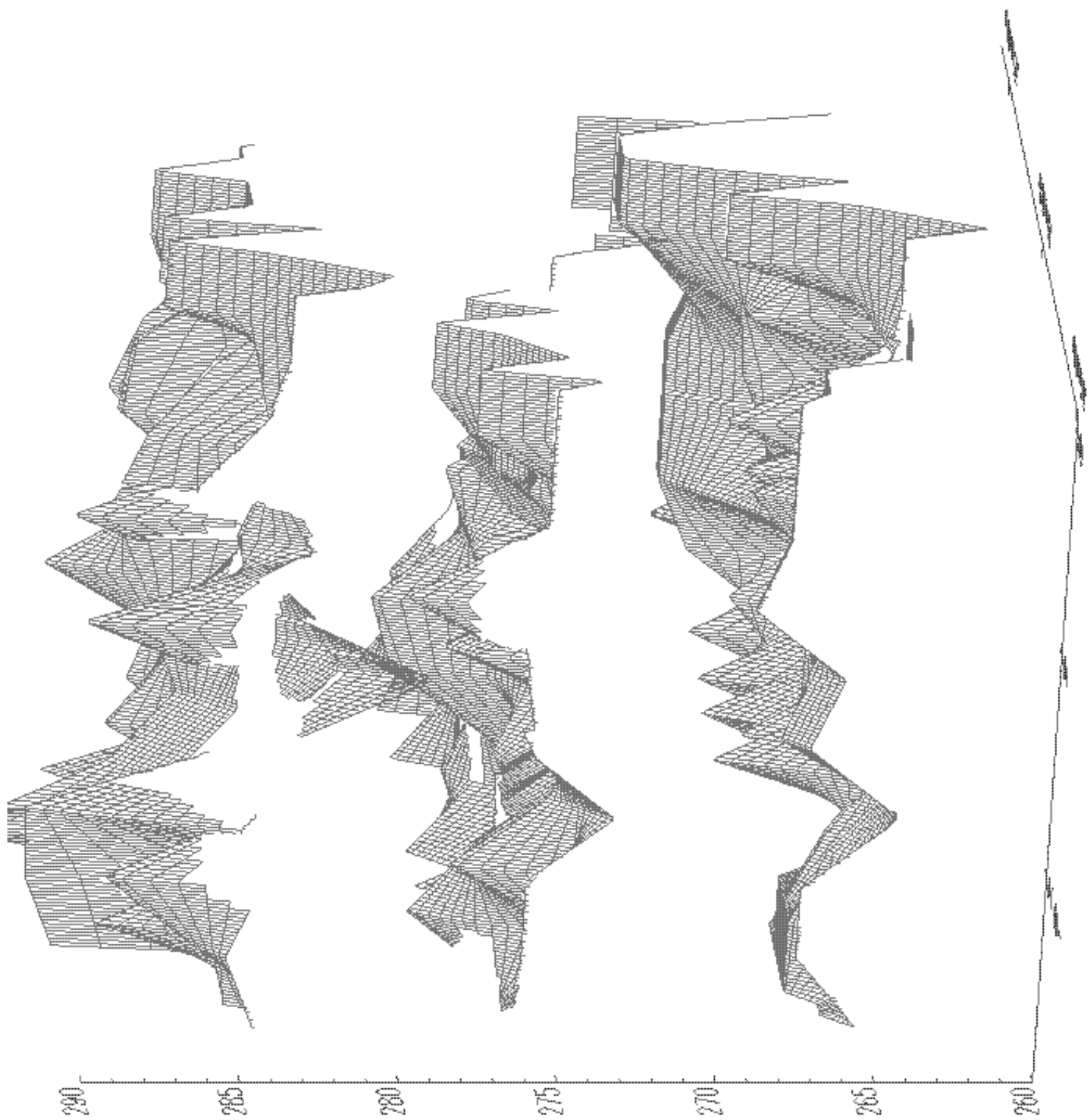
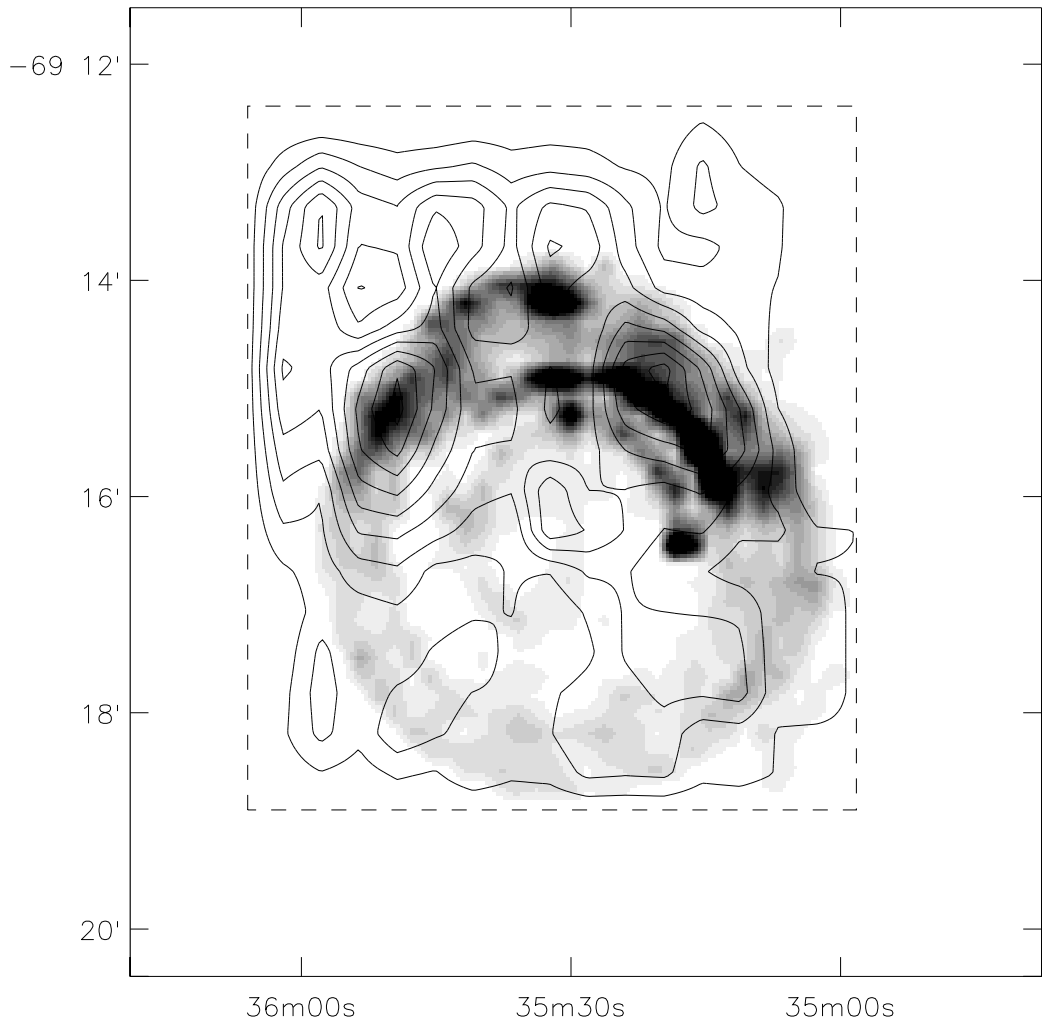
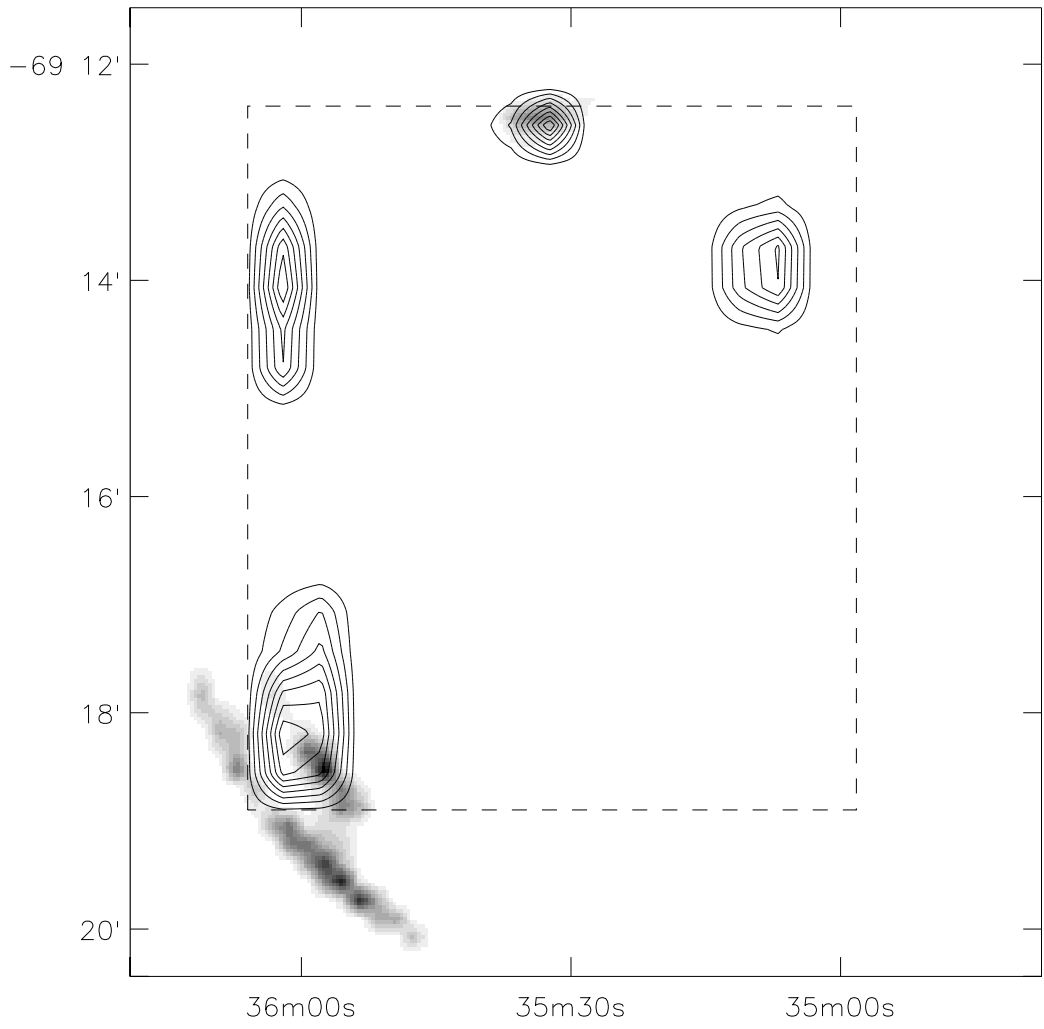
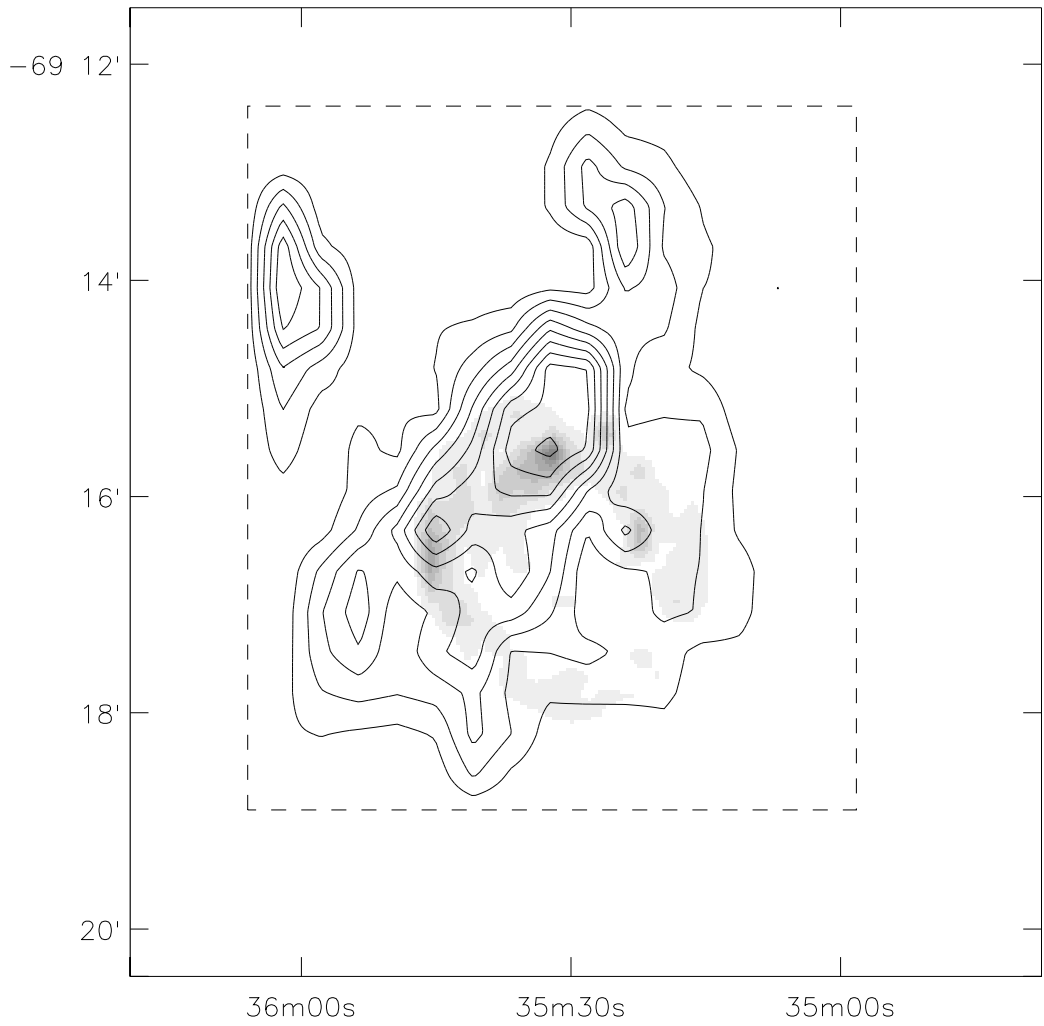
\end{document}